\documentclass[lettersize,journal]{IEEEtran}
\usepackage{amsmath,amsfonts}
\usepackage{algorithmic}
\usepackage{algorithm}
\usepackage{array}
\usepackage[caption=false,font=normalsize,labelfont=sf,textfont=sf]{subfig}
\usepackage{textcomp}
\usepackage{stfloats}
\usepackage{url}
\usepackage{verbatim}
\usepackage{graphicx}
\usepackage{cite}
\usepackage{upgreek}
\begin{document}

\title{Characterizing and modeling the influence of geometry on the performance of superconducting nanowire cryotrons}

\author{Alejandro Simon*, Reed Foster, Owen Medeiros, Matteo Castellani, Emma Batson, Karl K. Berggren, ~\IEEEmembership{Fellow,~IEEE}
        % <-this % stops a space
\thanks{*Corresponding author: alejansi@mit.edu}% <-this % stops a space
}

% ~\IEEEmembership{Member,~IEEE,}


\maketitle

\begin{abstract}
The scaling of superconducting nanowire-based devices to larger arrays is often limited by the cabling required to interface with each device. Cryogenic integrated circuits constructed from nanowire cryotrons, or nanocryotrons, can address this limitation by performing signal processing on chip. In this study, we characterize key performance metrics of the nanocryotron to elucidate its potential as a logical element in cryogenic integrated circuits and develop an electro-thermal model to connect material parameters with device performance. We find that the performance of the nanocryotron depends significantly on the device geometry, and trade-offs are associated with optimizing the gain, jitter, and energy dissipation. We demonstrate that nanocryotrons fabricated on niobium nitride can achieve a grey zone less than 210 nA wide for a 5 ns long input pulse corresponding to a maximum achievable gain of 48 dB, an energy dissipation of less than 20 aJ per operation, and a jitter of less than 60 ps. 
\end{abstract}

\begin{IEEEkeywords}
Superconductor, nanowire, cryotron, logic, amplification, comparator
\end{IEEEkeywords}

\section{Introduction}
\IEEEPARstart{S}{uperconducting} nanowire-based devices are a promising platform for computing, communication, and sensing technology \cite{hadfield_single-photon_2009, charaev_single-photon_2023, you_superconducting_2018, Toomey2020, sheinline_neuromorphic_2017, Lombo_2022}. To enable their use in a wider range of applications, scaling these devices to large arrays has attracted substantial interest in recent years \cite{castellani_nanocryotron_2024, oripov_superconducting_2023}. The largest arrays to date use delay-line readout to reduce cabling requirements \cite{oripov_superconducting_2023}, but the maximum count rate is intrinsically limited by the readout bus. Cryogenic integrated circuits could address this limitation by performing signal processing to reduce the amount of data sent off chip (and thus requiring fewer cables) while still preserving information of interest in high-count-rate environments.

\begin{figure*}
    \centering
    \includegraphics[width=0.95\textwidth]{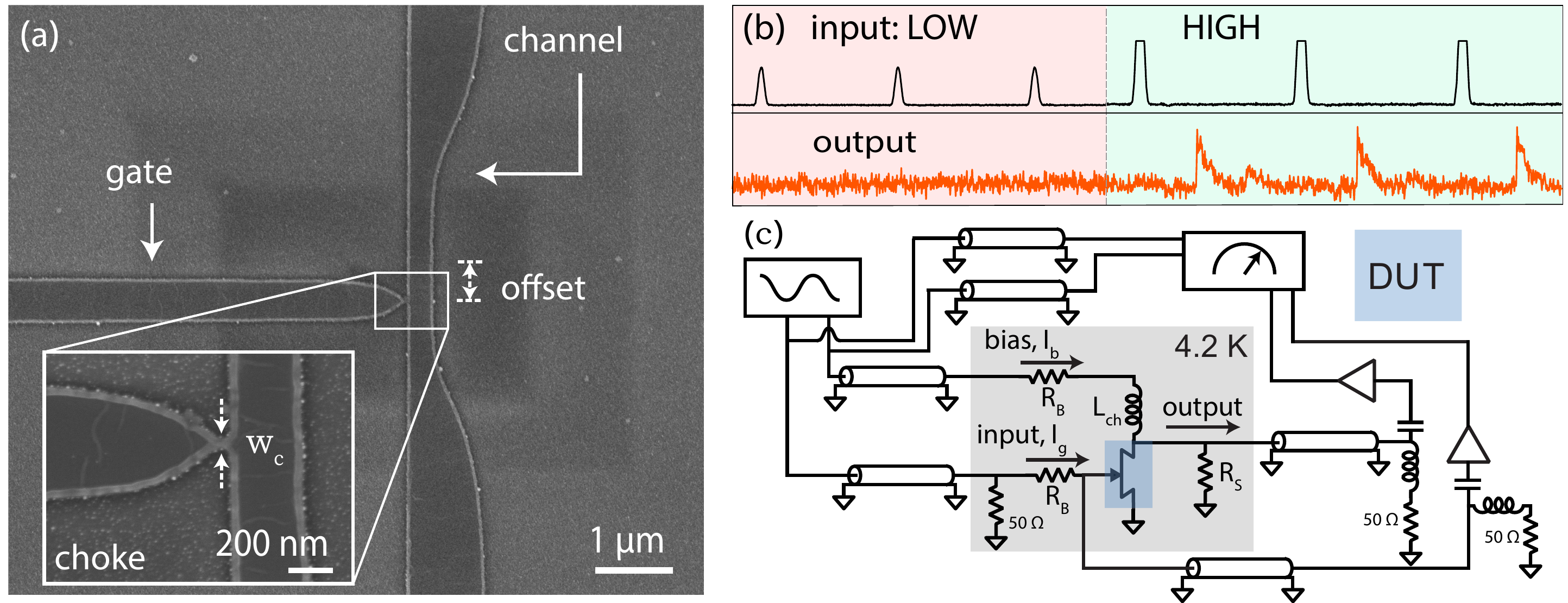}
    \caption{The nTron and electrical setup to test the device. (a) A scanning electron micrograph of an nTron with the components of the device labeled. The choke width, $w_{\mathrm{c}}$, and offset from the center of the channel were varied. Inset: A higher magnification micrograph of the nTron choke. (b) A time trace of the input and output voltages of the nTron during typical operation. A voltage pulse is recorded across the output only if the amplitude of the input pulse is above a switching threshold. (c) The electrical schematic used for the characterization of the nTron devices. Individual components of the schematic are described in the text and the device under test (DUT) is highlighted in blue. Measurements of the voltage across the terminals of the device allowed for precise determination of the device currents. When specified the voltage readout line connected to the gate was removed.}
    \label{fig:working-principle}
\end{figure*}

These circuits can be developed using the nanowire cryotron (nTron) \cite{mccaughan_superconducting-nanowire_2014, huang_monolithic_2024, zheng_characterize_2019}. The nTron is a three-terminal superconducting device that operates as a comparator, and is capable of amplifying signals and performing logical operations \cite{mccaughan_superconducting-nanowire_2014, buzzi_nanocryotron_2023, foster_shift_2023}. The device consists of two superconducting nanowires, the gate and channel, galvanically connected via a narrow constriction, the choke. The channel is wide relative to the choke, allowing it to support a larger supercurrent. A scanning electron micrograph of the nTron is depicted in Fig. \ref{fig:working-principle}a.

To operate the nTron, the channel is biased close to its critical current. Input is then supplied to the device via the gate. In the channel, the input and bias currents sum, and, if the input current is above the switching current of the choke, the choke switches to the normal state, injecting heat into the channel and suppressing its switching current. Together, the thermally-suppressed switching current and increased current density can switch the channel to the normal state, resulting in a large resistive domain (hotspot) forming and the bias current redirecting to the output. An example of the input and output signals of an nTron is shown in Fig. \ref{fig:working-principle}b for both low and high input current amplitudes. 

Though the integration of rapid superconducting flux quantum (RSFQ) and cryogenic complementary metal-oxide-semiconductor (cryo-CMOS) technology with superconducting nanowire devices has been demonstrated \cite{bardin_high-speed_2013, takeuchi_scalable_2020, hofherr_orthogonal_2012, yamashita_crosstalk-free_2012}, the nTron has several advantages that make it an attractive choice in many applications. In contrast to RSFQ devices, the nTron is easily monolithically integrated with other superconducting nanowire-based devices \cite{mccaughan_superconducting-nanowire_2014, huang_monolithic_2024, zheng_characterize_2019, zheng_superconducting_2020}, requiring no additional fabrication steps. The nTron can also operate unshielded in ambient magnetic fields up to $1\,\mathrm{T}$ \cite{draher_design_2023}, drive high-impedance loads \cite{zhao_nanocryotron_2017,mccaughan_superconducting-nanowire_2014}, and has a low jitter comparable to that of superconducting nanowire single-photon detectors (SNSPDs) \cite{zheng_characterize_2019, mccaughan_superconducting-nanowire_2014}. The nTron has no static power dissipation and its dynamic power consumption is low relative to cryo-CMOS. The nTron can also be made ultra-compact \cite{huang_monolithic_2024}.

To facilitate the use of nTrons for device readout and in cryogenic integrated circuits, it is necessary to characterize the nTron performance and understand how to optimize the device for desired characteristics \cite{mccaughan_superconducting-nanowire_2014, zheng_characterize_2019}. However, the relationship between important device parameters and the device geometry is not well understood. In this study, we characterize the nTron gain, grey zone, energy dissipation, and jitter and determine a relationship between these parameters and the device geometry. We also develop an electro-thermal model of the nTron, which provides deeper insight into the experimental results by relating material parameters to the device performance. 

\section{Methods}

\subsection{Fabrication}

To fabricate the nTron, $10\,\mathrm{nm}$ of niobium nitride ($\mathrm{NbN}$) was deposited onto a 4" silicon wafer with $300\,\mathrm{nm}$ thermal oxide using RF-biased magnetron sputtering. The sheet resistance of the film was $260\,\Omega/\ensuremath{\Box}$ at room temperature. Pads and alignment markers were defined with nLOF2035 resist using direct write photolithography and lift-off of $50\,\mathrm{nm}$ electron-beam evaporated gold with a $5\,\mathrm{nm}$ titanium adhesion layer. Nanowire devices were then patterned using ZEP530A as resist and transferred to the $\mathrm{NbN}$ via $\mathrm{CF}_4$/$\mathrm{Ar}$ reactive ion etching. A total of 448 devices were fabricated across seven chips. The width of the choke $w_{\mathrm{c}}$ was varied from $15\,\mathrm{nm}$ to $40\,\mathrm{nm}$ and the vertical offset of the choke was varied from the center of the channel to $3\,\upmu\mathrm{m}$ below the center. The channel and gate were $1\,\mathrm{\upmu m}$ wide and the constriction in the channel was $300\,\mathrm{nm}$ wide. The critical temperature of the patterned film was $8\,\mathrm{K}$. 

\subsection{Device characterization}
Fabricated devices were first screened using a room-temperature autoprober and a cryogenic probe station. The total device yield across the 448 devices was approximately 76\% and we anticipate this could have been improved with additional optimizations during the lithography steps. Device characterization was then performed on 27 devices across three of the fabricated chips in a liquid helium dewar at $4.2\,\mathrm{K}$ with the schematic depicted in Fig. \ref{fig:working-principle}c. Bias resistors were placed at cryo in series with the ports of the device with $R_{\mathrm{B}}=1\,\mathrm{k}\Omega$ unless otherwise specified, and a meandered nanowire with kinetic inductance $L_{\mathrm{ch}}\approx200\,\mathrm{nH}$ was included on chip in series with the channel. The inductor was bypassed by biasing the device through the output port for measurements where the voltage across the gate was recorded. $50\,\Omega$ resistors were placed in parallel with the bias resistors to prevent reflections, and a shunt resistor with $R_{\mathrm{S}}=5\,\Omega$ was placed in parallel with the output port of the nTron to prevent the device from latching into the normal state \cite{kerman_2009}. 

Direct current (DC) characterization was performed using a Stanford Research Systems SIM928 isolated voltage source to bias the device and a Keithley 2700 digital multimeter to record the output voltages. A room-temperature $1.9\,\mathrm{MHz}$ low-pass filter was placed in series with the input and output of the device to reduce noise. High-speed measurements were performed using a Keysight 33600a arbitrary waveform generator to bias the device and a LeCroy Waverunner 620Zi 20 GS/s high-speed oscilloscope to record the output voltage. The output was filtered with a mini-circuits ZFBT-4R2GW+ bias-tee and amplified with an RF-Bay LNA-2500 low-noise amplifier. 

The channel bias current $I_{\mathrm{b}}$ was set to approximately 90\% of the channel critical current, which varied between $55\,\upmu\mathrm{A}$ and $65\,\upmu\mathrm{A}$ at $4.2\,\mathrm{K}$ for the tested devices. Grey zone measurements were performed by applying a pulse of duration $\Delta t=5\,\mathrm{ns}$ to the gate and recording the probability of obtaining a switch in the channel. The resulting distribution was then fitted with an error function and the grey zone width for the channel and gate, $\delta I_{\mathrm{ch}}^{\mathrm{GZ}}$ and $\delta I_{\mathrm{g}}^{\mathrm{GZ}}$ respectively, was defined as the difference in gate currents $I_{\mathrm{g}}$ for which the fit was equal to 0.9 and 0.1. The energy dissipation of the device was determined by integrating the time trace of the product of the current into and the voltage across both the channel and gate. Jitter measurements were performed by recording a histogram of the delay time between applying a pulse with $\Delta t=5\,\mathrm{ns}$ to the gate with $I_{\mathrm{g}}=40\,\upmu\mathrm{A}$ and the resulting switch in the channel for 10,000 input pulses. The jitter was defined as the root mean squared deviation of the distribution of delay times.

\subsection{Electro-thermal model}
The effective electron and phonon temperatures of the nTron, $T_e$ and $T_{\mathrm{ph}}$ respectively, can be modeled using coupled two-temperature heat equations \begin{equation}
    C_{e}(T_e) \frac{d T_e}{d t} = - \frac{C_{e}(T_e)}{\tau_{e-\mathrm{ph}}} (T_e - T_{\mathrm{ph}}) + \nabla\kappa_e(T_e)\nabla T_e + \Vec{j} \cdot \Vec{E}
\end{equation}
\begin{equation}
\begin{aligned}
     C_{\mathrm{ph}}(T_{\mathrm{ph}})& \frac{d T_{\mathrm{ph}}}{d t} = \frac{C_{\mathrm{ph}}(T_{\mathrm{ph}})}{\tau_{\mathrm{ph-}e}} (T_e - T_{\mathrm{ph}}) \\ & - \frac{C_{\mathrm{ph}}(T_{\mathrm{ph}})}{\tau_{\mathrm{esc}}} (T_{\mathrm{ph}} - T_{\mathrm{b}}) + \nabla\kappa_{\mathrm{ph}}(T_{\mathrm{ph}})\nabla T_{\mathrm{ph}},
\end{aligned}
\end{equation} where $T_{\mathrm{b}}$ is the bath temperature, $C_e(T_e)$ is the electron heat capacity determined by the BCS result for the superconducting state \cite{bardeen_theory_1957} and $C_e(T_e)=\gamma T_e$ for the normal state \cite{kittel_introduction_2004}, $C_{\mathrm{ph}}(T_{\mathrm{ph}})=\alpha T_{\mathrm{ph}}^3$ is the phonon heat capacity, $\tau_{e-\mathrm{ph}}=\tau_0 T_{e}^{-3}$ is the electron-phonon time constant, $\tau_{\mathrm{ph}-e}$ is the phonon-electron time constant, $\tau_{\mathrm{esc}}$ is the phonon escape to substrate time constant, $\kappa_e(T_e)$ is the electron thermal conductivity modeled using the Wiedemann-Franz relation \cite{kittel_introduction_2004}, $\kappa_{\mathrm{ph}}(T_{\mathrm{ph}})=\alpha D_{\mathrm{ph}} T_{\mathrm{ph}}^3$ is the phonon thermal conductivity, $\Vec{j}$ is the current density, and $\Vec{E}$ is the electric field \cite{baeva_thermal_2018, perrin_dynamic_1987, baghdadi_multilayered_2020}. We also make use of the equilibrium relation $C_e/\tau_{e-\mathrm{ph}}=C_{\mathrm{ph}}/\tau_{\mathrm{ph}-e}$. Typical material parameters for $\mathrm{NbN}$ were used, with $\gamma = 240\,\mathrm{J/m^3 K^2}$, $\alpha=1.03\,\mathrm{J/m^3K^4}$, $\tau_0\approx10\,\mathrm{ps}$, $\tau_{\mathrm{esc}}\approx 40\,\mathrm{ps}$, and $D_{\mathrm{ph}}=0.33\,\mathrm{cm}^2/\mathrm{s}$ \cite{baeva_thermal_2018, baghdadi_multilayered_2020}. The coupled heat equations were solved for each nTron geometry and test circuit using a finite element simulation in the COMSOL Multiphysics Solver software.

\section{Results}
\subsection{DC Characterization}
\begin{figure}
    \centering
    \includegraphics[width=0.95\linewidth]{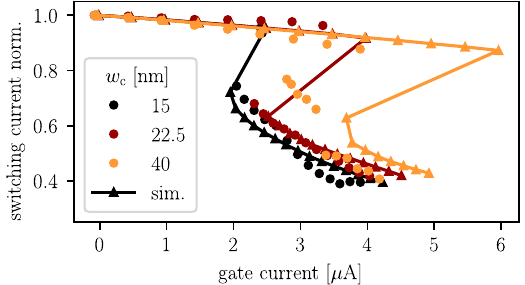}
    \caption{The channel switching current normalized to the channel critical current as a function of the measured gate current for nTrons with a $15\,\mathrm{nm}$, $22.5\,\mathrm{nm}$, and $40\,\mathrm{nm}$ choke width. There is a significant suppression of the channel switching current as the gate current increases. When the gate current exceeds the choke switching current, the choke becomes resistive leading to a decrease in the actual current delivered to the gate. The results from the electro-thermal model simulations are displayed with the experimental data.}
    \label{fig:DC}
\end{figure}

As depicted in Fig. \ref{fig:DC}, the channel switching current $I_{\mathrm{ch}}^{\mathrm{sw}}$ is suppressed as $I_{\mathrm{g}}$ increases. For $I_{\mathrm{g}}$ less than the switching current of the choke $I_{\mathrm{g}}^{\mathrm{sw}}$, the suppression is linear due to the summation of currents. When $I_{\mathrm{g}} > I_{\mathrm{g}}^{\mathrm{sw}}$, the choke switches to the normal state, injecting heat into the channel and significantly reducing $I_{\mathrm{ch}}^{\mathrm{sw}}$. Since the choke becomes resistive, the actual current delivered to the gate is also reduced. When the choke's critical current is larger than the difference between the channel critical current and $I_{\mathrm{b}}$, the increase in the channel current from $I_{\mathrm{g}}$ can be sufficient to switch the channel without switching the choke. This effect is visible in Fig. \ref{fig:DC} for the device with $w_{\mathrm{c}} = 40\,\mathrm{nm}$, where a significant suppression of the channel critical current occurs at $I_{\mathrm{g}}\approx4\,\upmu\mathrm{A}$ and $I_{\mathrm{g}}\approx3\,\upmu\mathrm{A}$, corresponding to the channel switching before and after the choke switches respectively.

The results of a finite element simulation using the electro-thermal model for the nTron are also displayed in Fig. \ref{fig:DC}. There is qualitative agreement between the theoretical and experimental curves, suggesting that the operation of the device is well-explained by electro-thermal physics. Quantitative disagreement between the theory and experiment is likely due to differences in the values of the material parameters from the literature and the tested films, defects in the material, and, likely particularly important for wider $w_{\mathrm{c}}$, quantum and thermal fluctuations that are not accounted for in the electro-thermal model.

\subsection{Grey zone \& gain}
\begin{figure}
    \centering
    \includegraphics[width=0.95\linewidth]{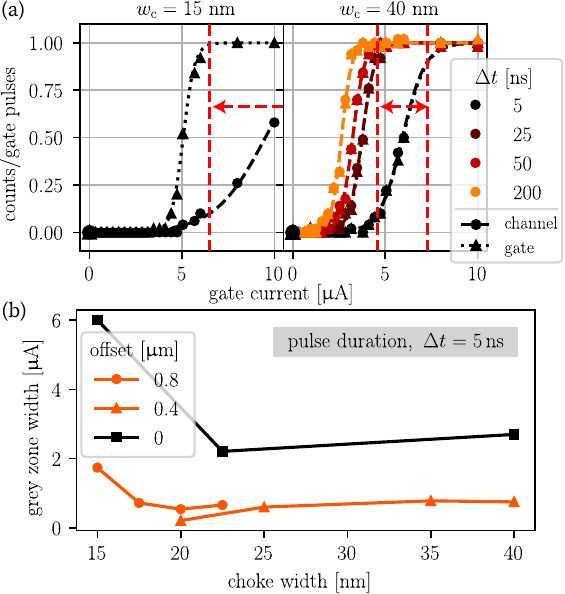}
\caption{Measurements of the device grey zone. (a) Plots of the number of counts, or switches in the channel and gate, per gate pulse for the $15\,\mathrm{nm}$ and $40\,\mathrm{nm}$ choke widths for different gate currents. An error function is fit to each switching distribution, and the grey zone is defined as the range of gate currents for which the fitted error function is between 0.1 and 0.9 as indicated by the dashed red lines. For clarity, only data for the $5\,\mathrm{ns}$ long pulse is shown for the $15\,\mathrm{nm}$ wide choke. (b) The channel grey zone width as a function of the choke width for a $5\,\mathrm{ns}$ pulse duration. The orange denotes devices with a gate offset and measured with the gate not shunted by the gate voltage readout. The $0.4\,\upmu\mathrm{m}$ offset devices were biased with $R_{\mathrm{B}}=10\,\mathrm{k}\Omega$.}
    \label{fig:grey-zone}
\end{figure}

As shown in Fig. \ref{fig:grey-zone}a, $\delta I_{\mathrm{ch}}^{\mathrm{GZ}}$ and $\delta I_{\mathrm{g}}^{\mathrm{GZ}}$ decrease with increasing $\Delta t$. Physically, this corresponds to an increase in the injected heat and additional time for a quantum or thermal fluctuation to occur in the channel (or gate) that switches it to the normal state. We expect this process to be well-described by an Arrhenius rate equation \cite{salvoni_activation_2022}. 

In Fig. \ref{fig:grey-zone}b, a minimum $\delta I_{\mathrm{ch}}^{\mathrm{GZ}}=210\,\mathrm{nA}$ occurs when $w_{\mathrm{c}}=20\,\mathrm{nm}$, and $\delta I_{\mathrm{ch}}^{\mathrm{GZ}}$ increases substantially for devices with a narrower $w_{\mathrm{c}}$, with $\delta I_{\mathrm{g}}^{\mathrm{GZ}}$ and $\delta I_{\mathrm{ch}}^{\mathrm{GZ}}$ taking on different values as shown in Fig. \ref{fig:grey-zone}a. This effect was consistent across the four devices tested with $w_{\mathrm{c}}=15\,\mathrm{nm}$ for $\Delta t=5\,\mathrm{ns}$. We do not conclude a physical origin for the dependence of $\delta I_{\mathrm{ch}}^{\mathrm{GZ}}$ on $w_{\mathrm{c}}$; however, as will be described in Section \ref{sec:energy}, the energy dissipation in the choke did not differ significantly between devices with $w_{\mathrm{c}}=15\,\mathrm{nm}$ and $w_{\mathrm{c}}=22.5\,\mathrm{nm}$, which suggests the dependence is not due to a reduction in the generated heat. In measurements that did not include a connection from the gate to the $50\,\Omega$ readout, the same trends were observed, along with a reduction in $\delta I_{\mathrm{ch}}^{\mathrm{GZ}}$ that occurred since the choke was no longer shunted by a $50\,\Omega$ line.

An upper-bound on the device gain can be estimated as the ratio of $I_{\mathrm{b}}$, i.e., the output current, to $\delta I_{\mathrm{ch}}^{\mathrm{GZ}}$, i.e. the minimum detectable input signal level. The largest achievable gain was observed in the device with $w_{\mathrm{c}} = 20\,\mathrm{nm}$ driven by a $10\,\mathrm{k}\Omega$ resistor. This device had a grey zone of $210\,\mathrm{nA}$ with $I_{\mathrm{b}}=53\,\upmu\mathrm{A}$, resulting in a maximum gain of $48\,\mathrm{dB}$. 

\subsection{Energy dissipation}
\label{sec:energy}
The energy dissipation per switching event of the nTron was found to range between $20\,\mathrm{aJ}$ and $30\,\mathrm{aJ}$ for $\Delta t=5\,\mathrm{ns}$. This can be reduced by decreasing the film thickness or designing a narrower channel to reduce $I_{\mathrm{b}}$, however, this reduces the output current of the device. $R_{\mathrm{S}}$ could also be decreased at the cost of additional reset time and decreased output current. We do not measure the operating speed of the nTron; however, other work has operated the device at $615.4\,\mathrm{MHz}$ \cite{zheng_characterize_2019}.

\begin{figure}
    \centering
    \includegraphics[width=0.95\linewidth]{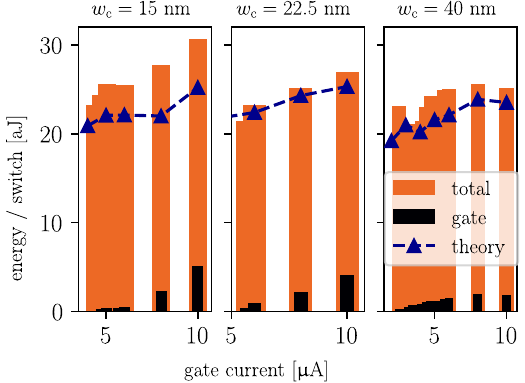}
    \caption{Total energy dissipation and the contribution from the gate as a function of the applied gate current for devices with a $15\,\mathrm{nm}$, $22.5\,\mathrm{nm}$, and $40\,\mathrm{nm}$ choke width. The channel bias current for these devices was $52\,\upmu\mathrm{A}$, $50.5\,\upmu\mathrm{A}$, and $50.5\,\upmu\mathrm{A}$ respectively. The energy dissipation calculated from the electro-thermal model is displayed alongside the experimental data.}
    \label{fig:energy}
\end{figure}

In Fig. \ref{fig:energy} it is shown that the energy dissipated in the gate for $w_{\mathrm{c}}=15\,\mathrm{nm}$ and $w_{\mathrm{c}}=22.5\,\mathrm{nm}$ exhibits a quadratic dependence on the gate current. In these cases, the choke initially switches and suppresses $I_{\mathrm{ch}}^{\mathrm{sw}}$ via Joule heating in the choke. On the other hand, the device with $w_{\mathrm{c}}=40\,\mathrm{nm}$ can support enough supercurrent to increase the current density in the channel beyond the critical current density before the choke switches. Because the channel switches before the choke, there is a weak dependence of the energy dissipation in the gate on the gate current and a slight reduction in the total energy dissipation of the device.

The $\mathrm{aJ}$-scale energy dissipation permits integration of many devices onto a single chip: assuming an activity factor of 10\%, and a switching speed of $200\,\mathrm{MHz}$, the power dissipation for ten million devices would be well below $10\,\mathrm{mW}$ at $4\,\mathrm{K}$. When considering applications such as superconducting detector readout, where the detectors are already cold, the cooling penalty is somewhat irrelevant, and the power dissipation of these devices is amenable to large-scale circuits that could be monolithically integrated with detectors. 

\subsection{Device jitter}
\begin{figure}
    \centering
    \includegraphics[width=0.95\linewidth]{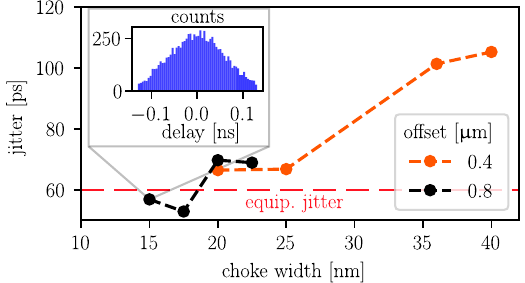}
    \caption{The nTron jitter measured as a function of the choke width for the $0.4\,\upmu\mathrm{m}$ and $0.8\,\upmu\mathrm{m}$ devices. The jitter decreases as the choke width is reduced. Inset: The distribution of delay times for the nTron with a $\mathrm{15\,nm}$ wide choke.}
    \label{fig:jitter}
\end{figure}

As depicted in Fig. \ref{fig:jitter}, the nTron jitter decreases with decreasing $w_{\mathrm{c}}$. The minimum jitter measured was limited by the jitter of the measurement equipment, which was $60\,\mathrm{ps}$, causing the observed trend in the jitter to become less pronounced as the measurement approached $60\,\mathrm{ps}$. Due to this limitation, we place an upper bound of $60\,\mathrm{ps}$ on the minimum jitter of these devices, which is comparable to the typical jitter of a large-area SNSPD.

\section{Conclusion}
We demonstrate that the nanowire cryotron is well-described by an electro-thermal model and by designing the device appropriately, the nanowire cryotron can achieve a grey zone width down to $210\,\mathrm{nA}$ for a $5\,\mathrm{ns}$ long pulse, energy dissipation of $20~\mathrm{aJ}$ per operation, and sub-$60\,\mathrm{ps}$ jitter. These performance metrics make the device a promising platform for developing large-scale cryogenic integrated circuits for interfacing with other nanowire devices. Based on measurements in this study, we estimate that these circuits could easily scale to tens of millions of devices on a single chip. Future work can extend this model and the experimental results shown here to engineer the performance of the nTron for a wider range of applications. 

\section*{Acknowledgments}
This work was funded by the Department of Energy Office of Science Research Program for Microelectronics Codesign (LAB 21-2491). A.S. acknowledges support from the NSF GRFP and MIT Vanu Bose Presidential fellowship programs. O.M. acknowledges support from the NDSEG fellowship program. We thank the MIT.nano staff for their assistance and guidance in fabricating the devices presented in this work.

\vfill

\end{document}